\documentclass[12pt,a4paper]{extarticle}

\usepackage{ifpdf}
\pdfoutput=1 
\usepackage[utf8]{inputenc}
\usepackage{amsmath}
\usepackage{amsfonts}
\usepackage{amssymb}
\usepackage{caption}
\usepackage{subcaption}
\usepackage{graphicx}
\usepackage[bookmarks=false]{hyperref}
\usepackage{authblk}
\usepackage[margin=1in]{geometry}
\usepackage[english]{babel}
\usepackage{authblk}
\usepackage{lineno}
\usepackage{xmpmulti}
\usepackage{animate}
\usepackage{pslatex} 
\usepackage{threeparttable}

\everymath=\expandafter{\the\everymath\displaystyle}

\begin{document}

\title{Robustness of urban road networks based on spatial topological patterns} 
\author[1]{Felipe Vaca-Ramírez}
\affil{NRU Higher School of Economics}
\date{}
\maketitle

\begin{abstract}
\medskip
During the last decade, road network vulnerability assessment has received an increasing attention. On one hand, it is due to the significant advances in Network Science and the potentialities that its tools offer. On the other hand, it is due to its utility for urban planning and emergency response. Despite these facts and the increasingly available data, related work is still sparse in Latin America, even more so in Ecuador. Due to its geographical, historical, and social characteristics, the city of Quito is considered as a case study. At first, the spatial distributions of several topological centrality measures are analyzed. As expected, there are hotspots where high values of these measures concentrate. These results serve to further simulate several strategies for disconnecting the urban road network. Finally, we observe that centrality-based strategies are more effective than randomly-based strategies in disconnecting the network.

\end{abstract}

\section{Introduction}

Network analysis has been an increasingly used approach for studying the structure of cities. It usually consists on representing the urban street network as a graph and analyzing its topological properties. Among these, centrality measures (e.g., degree, closeness, betweenness) are the most largely used in the related literature (e.g., see \cite{crucitti2006centrality}, \cite{courtat2011mathematics} \cite{zhao2016network} \cite{sienkiewicz2005statistical}). 

It is worth to note that cities are constrained by geographical conditions, and therefore, their urban networks are spatially embedded. In the literature they are consider as a special case of spatial networks. This spatial constrain prevents from the immediate application of centrality measures, which were usually developed for social networks, to spatial networks. Indeed, the spatial embedding makes a city behave neither like a classical scale-free network nor a small-world \cite{BOCCALETTI2006175}. For this reason, some of the properties of spatial networks -e.g. degree- become trivial, while others -e.g. betweenness centrality- become more important \cite{barthelemy2011spatial}. 

Unfortunately, most of the studies that attempted to characterize the urban structure have focused on the topological aspect, and have put aside the spatial aspect. Furthermore, when the spatial aspect of topological centrality measures has been taken in account, studies have usually limited themselves to partially plot the values of such measures on a map \cite{crucitti2006centrality}.

Another aspect that differentiates spatial networks from other types of networks is that the network display is not arbitrary in the first case. Furthermore, if the nodes of urban road networks are not randomly located on the space, then there is no reason to assume that the values of topological centrality measures are randomly distributed over such space. When spatial randomness does not hold, spatial patterns of such topological measures (e.g. hotspots) may emerge. Discovering these patterns may be of interest of urban planners because they may reveal potential relations between such patterns and mobility, performance of transportation systems, land use, and urban form \cite{xie2007measuring}.

According to \cite{barthelemy2011spatial}, although some of these ideas had been already discussed at least 40 years ago, the growing availability of data, computation power, and knowledge complex networks has helped to gain new insights on this topic. Recently, some researchers have devoted their efforts towards studying the spatial distribution of topological measures based on kernel density estimations \cite{liu2015road}, using percolation models \cite{arcaute2015hierarchical} \cite{li2015percolation} in urban networks, and assessing the network robustness by means of simulations \cite{duan2013structural}.

This project focuses in the latter issue since structural robustness analysis can describe the survivability of a city road network that is under attack and can help improve functions such as urban planning and emergency response \cite{khademi2015transportation}.

In the literature, such survivability is frequently described by a reference to network vulnerability, instead of by a direct reference to network robustness. Although several definitions of vulnerability have been proposed, a highly cited one corresponds to \cite{berdica2002introduction}: ``Vulnerability in the road transportation system is a susceptibility to incidents that can result in considerable reductions in road network serviceability''.

\cite{mattsson2015vulnerability} does a comprehensive review of approaches to network vulnerability by identifying two major approaches: one with roots in graph theory which studies the vulnerability of transport networks based on their topological properties; and other that also represents the demand and supply side of the transport systems. The studies from the first approach usually start computing centrality measures (sometimes at different levels). Then  they simulate attacks by removing nodes or edges according to some criteria, such as centrality indices. Finally, they analyze the changes in some topological properties of the biggest connected component \cite{duan2013structural} \cite{berche2012transportation} \cite{latora2005vulnerability} or increase in user travel cost/time \cite{rupi2015evaluation}.

A concept closely related with node/link failures in road networks is accessibility. It is expected that disruption in the network reduce the accessibility to some areas. Some authors have paid attention to this topic and combined these ideas with natural and social aspects. For example, \cite{nyberg2013indicators} assessed the vulnerability of a Swedish road network by performing disruptions (which represent tree falls on roads produced by wind storms) and analyze the accessibility that rescue services have to zones where elderly people live. \cite{novak2014link} developed a new measure of accessibility to emergency service facilities (ESFs) called critical closeness accessibility (CCA). It is based on the idea of closeness centrality, and it is used to perform link removals in the network and study how network disruptions may deprive certain segments of the population of the ability to reach some set of ESFs. \cite{bono2011network} study the accessibility reduction that emergency service teams may have experienced during the 2010 Haiti earthquake by representing the urban environment as a grid, setting links between a cell and its neighbouring cells, and introducing a cost of of reaching each cell.

Most studies have shown that transport networks are more robust to random attacks than to directed attacks (e.g. a sequential removal nodes or edges with higher centrality). Nevertheless, the role of space and economic equity in network vulnerability are yet aspects to be studied. The first one has to be with how the location of important nodes or edges, or more generally, the spatial structure of the network, influences the impact of a disruption in the network \cite{grubesic2008comparative}. The second one has to be with which social or economic classes are more affected when network disruptions occur \cite{mattsson2015vulnerability}. 

We claim that these aspects are relevant in Latin America since class relations may have influenced the urban growth and sprawl of the city. According to \cite{portes2005free}, in the beginning, elites and middle classes were located in the city center. From 1930 to 1970 the process a industrialization occurred in the continent. It increased labor demand of cities and caused a massive internal migration process. Due to the population growth, housing demand increased and house prices raised above wages. Without public policies to provide housing solutions, the working classes were forced to create settlements on the cities periphery\cite{barros2005simulating}, most of the time without regulation. This process was accompanied by the movement of elites and middle classes from the city center to increasingly remote suburban areas, away from those occupied by the poor.

In this context, this project will explore some street network patterns that emerge in a city with the previously mentioned characteristics. The city of Quito, the capital of Ecuador, is considered as a case study. It contains agglomeration of facilities, high use of access roads, and a displacement of the elites from the center to the North and Valleys \cite{ryder2004land} \cite{bustamante2014crecimiento}. Furthermore, four zones of interest within it (the North, the South, Los Chillos Valley, and Tumbaco-Cumbayá Valley) are also explored to show some spatial patterns.

There are other reasons that make Quito a focus of interest. First, it has an elongated shape which heavily constraints the alternatives to move from one place to another. Second, some of its zones are within risk zones under volcano eruption \cite{Aguilera2004}. Third, some streets have been built over concrete box tunnels, which in turn rerouted rivers. Sinkholes may occur because of the accumulation of trash and heavy rain (as the event of 2008 described in \cite{salazar2009trebol}). Fourth, it is located in one of the most exposed zones to seismic hazard at the national scale \cite{miduvi2014norma}. Then damages to elevated road structures can occur in case of an earthquake, and therefore road disruptions \cite{demoraes2005seismic}. These and other risks are comprehensively described in \cite{demoraes2015movilidad}. Hence, such document will serve as a ground truth for testing if the topological structure of the urban road network can give account of vulnerable points of the city.

The first step of our methodology consists on computing several centrality measures of nodes (one of them is based on bus routes) and observe the spatial distribution of important nodes in the study area. Then we simulate directed attacks to most central nodes and compare their effects in the robustness of the network with the effects of random attacks. These ideas will be extended in section \ref{sec.method}.

The corresponding results will be further described in section \ref{sec.data}. Analogously to the existing literature, we expect that the failure (or extraction) of a subset of the most central nodes from the network be more harmful than a random extraction of nodes for the connectivity of such network. Furthermore, we also expect important differences between the robustness of several zones of a city, i.e., there are parts of the city that become disconnected faster than others when performing attacks, since urban street networks are spatially embedded. 

Finally, we think that this project may offer insights about the role of highly central nodes, usually spatially clustered in urban road networks, and serve as a tool for urban planning. Nevertheless, there are some limitation in our analysis that should be taken in account. They will be presented in section \ref{sec.Conclusions}.

\section{Methodological Background}\label{sec.method}

A graph or network G is a set of vertices or nodes V connected by edges E. The edges can have direction in which case, G is called a directed graph. Furthermore, edges can also have weights (usually, a real number), which can represent intensity, flow or frequency of the relation between two nodes. In this case, G is called weighted graph.

We have considered some of the indices that according to \cite{barthelemy2011spatial} may be useful to characterize spatial networks.

\subsection{Centrality}

Centrality refers to the importance of a node in the network. A variety of measures have been proposed to quantify centrality, and the differences can be accounted for what each context understands as ``important''. According to the following indices, greater values of an index implies that the node is more central. 

\subsubsection{Degree Centrality}
The degree of a node is the number of edges connected to it. In weighted graphs, the degree of a node is the sum of the weights of the adjacent edges to the node. If the graph is directed, the nodes have an in-degree (number of incoming edges) and and out-degree (number of outcoming edges). The values of this index can be normalized by 1/(n-1), where n is the number of nodes of the network.

\subsubsection{Closeness Centrality}
This index measures how close is a node to all other nodes in the network, considering shortest paths. For a given node i, closeness centrality is defined as the reciprocal of the average distance from i to the rest of nodes of the network:

\begin{equation}
C_{C}(i)=\frac{1}{\sum_{j \in V}d_{ij}}
\end{equation}

where $d_{ij}$ is the distance between nodes i and j. The values of this index can be normalized by 1/(n-1), where n is the number of nodes of the network.

\subsubsection{Betweenness Centrality}
This index measures how often a node belongs to the shortest paths of the other vertices. For a given node i, betweenness centrality is defined as the sum of the fractions of shortest paths between nodes that cross node i:

\begin{equation}
C_{B}(i)=\sum_{j,k\in V} \frac{\sigma(j,k|i)}{\sigma(j,k)}
\end{equation}

where V is the set of nodes, $\sigma(j,k)$ is the number of shortest paths between j and k, and $\sigma(j,k|i)$ is the number of such paths that cross the node i different from j and k. If j = k, then $\sigma(j,k)=1$, and if $i \in {j,k}$, then $\sigma(j,k|i)=0$ . The values of this index can be normalized by $2/((n - 1)(n - 2))$ for non directed graphs and by $1/((n - 1)(n - 2))$ for directed graphs, where n is the number of nodes in G ~\cite{brandes2008variants}. This measure is considered as relevant since the removal of those nodes with highest betweenness will mostly affect the flows in the network ~\cite{newman2010networks}.

\subsubsection{Load Centrality}
This index is a slight variation of betweenness centrality. In particular, for a given node, load centrality is the fraction of all shortest paths that pass through that node ~\cite{goh2001universal}.

\subsection{Kernel Density Estimation for Points}
Spatial patterns (such as hotspots) may emerge when points are not uniformly distributed in space. The detection of these patterns can give insights into spatial inequality and spatial autocorrelation.

The density of points gives and indication of its spatial distribution. A non-parametric estimation of such density can be done by means of kernel smoothing,  which for each point, can be seen as a weighted average of the values of its nearest neighbours. 

In the rest of this subsection we will briefly introduce this approach. We will adopt the definitions from \cite{silverman1986} and assume that all points are in the 2-dimensional space.

Let $X_1$, $X_2$,...,$X_n$ be 2-dimensional points. The corresponding kernel density estimator with kernel $K$ and bandwidth $h$ is defined by
\begin{equation}
    \hat{f}(x) = \frac{1}{nd^2}\sum_{i=1}^n K \left( \frac{x-X_i}{h} \right)
\end{equation}

where $K(x)$ is a function defined for a 2-dimensional point $x$, and satisfying the following properties: 

$K(x) \geq 0$,  for all x,
\\
$\int_{R^2}K(x)dx = 1$.

There are several kernels, but in this case, we will consider the quartic kernel function, which is said to be useful in the 2-dimensional space: 

    \[K_2(x) = \begin{cases} 
       3 \pi^{-1}(1-x^T x)^2 & if \quad x^T x < 1 \\
       0 & otherwise
   \end{cases}
\]

The are also several criteria for setting the bandwidth ($h$), and here we will set it considering mobility patterns, i.e. 1000 meters.

\subsection{Community Structure Detection}

Community detection methods may show vulnerabilities of urban networks \cite{duan2013structural}. In principle, these methods try to find groups of nodes which exhibit high connectivity within groups and low connectivity between groups. 
In this project, we have considered one of the simplest community detection approaches. It consists on sequentially removing the nodes with higher centrality (or randomly chosen) from the graph and observing how the biggest connected component properties (e.g., its size) change as a result. Collectively, these approaches are described under the name of divisive algorithms in \cite{fortunato2010community}. Unlike required by the Girvan Newman algorithm \cite{newman2004finding}, centralities are not recomputed after each node removal, because of the size of our graphs and computation time.

\section{Data and Results}\label{sec.data}

\subsection{Data}\label{sec.data_des}

We have considered the road network of Quito as a primal graph, i.e., the nodes are the intersections between streets and the edges are the links between them. It is a connected, directed and weighted graph, where weights are the lengths of street segments. The subgraphs corresponding to the North, South, Los Chillos Valley, and Tumbaco-Cumbayá Valley  preserve the same characteristics. The computation of network indices was made on Python with \textit{NetworkX} package \cite{hagberg-2008-exploring}.

Additional to topological information we have also included information on mobility patterns. In particular, we have considered, as a measure of node importance, the number of times a bus route crosses for each node. This data is only available for the northern and southern parts of the city.

\begin{figure}[!h]
	 \centering   	        
   \begin{subfigure}[t]{0.3\textwidth}
   \centering 
        \includegraphics[scale=0.475]{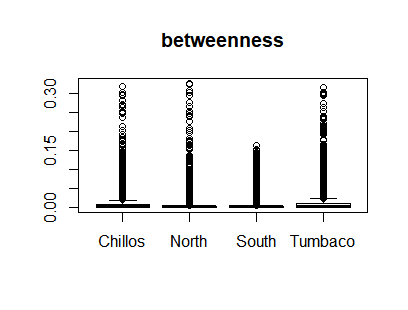}
        \caption{}
    \end{subfigure}
    ~ \hfill   
   \begin{subfigure}[t]{0.3\textwidth}
   \centering 
        \includegraphics[scale=0.475]{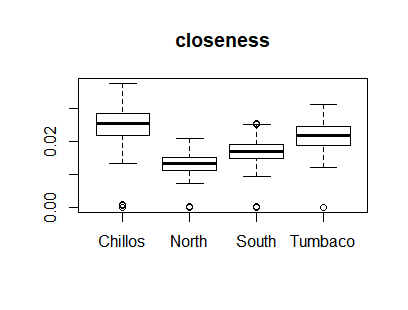}
        \caption{}
    \end{subfigure}
    ~ \hfill 
   \begin{subfigure}[t]{0.3\textwidth}
   \centering 
        \includegraphics[scale=0.475]{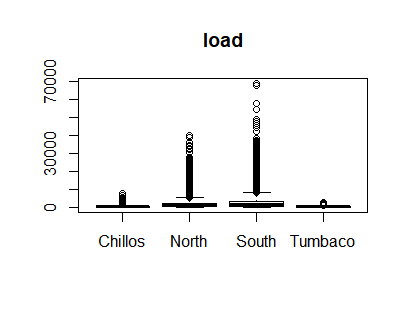}
        \caption{}
    \end{subfigure}
    ~ \hfill  
   \begin{subfigure}[t]{0.3\textwidth}
   \centering 
        \includegraphics[scale=0.475]{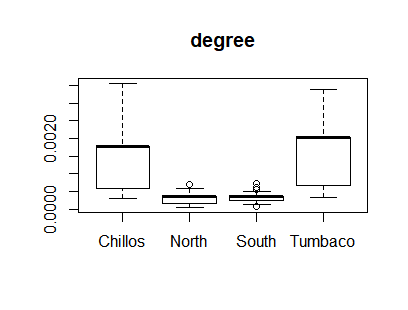}
        \caption{}
    \end{subfigure}
    ~ \hfill 
   \begin{subfigure}[t]{0.3\textwidth}
   \centering 
        \includegraphics[scale=0.475]{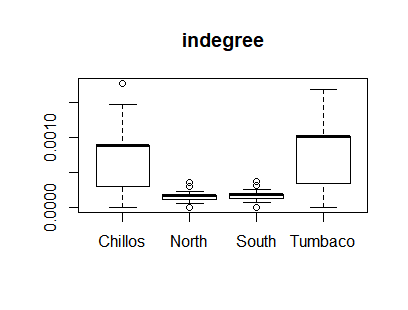}
        \caption{}
    \end{subfigure}
    ~ \hfill 
   \begin{subfigure}[t]{0.3\textwidth}
   \centering 
        \includegraphics[scale=0.475]{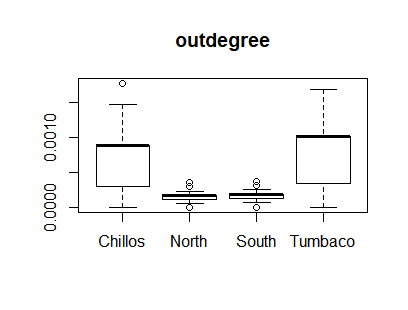}
        \caption{}
    \end{subfigure}
    ~ \hfill     
   \begin{subfigure}[t]{0.3\textwidth}
   \centering 
        \includegraphics[scale=0.475]{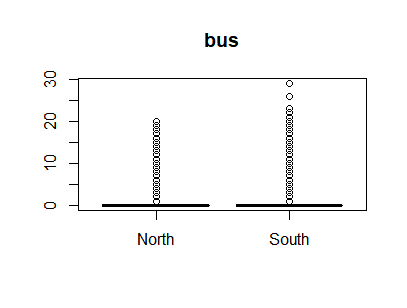}
        \caption{}
    \end{subfigure}
    \caption{Boxplots of centrality indices}  
\label{boxplots}
\end{figure}

Table \ref{sumstat} and Figure \ref{boxplots} present some characteristics of data and suggest some differences between the structure of zones. 

Hereinafter, we will compare the North with the South, and the valleys among them due to their similarities in subgraph sizes and antiquity of settlements (the North and South were built way before the valleys).

From figure \ref{boxplots}, it is clear that distributions deviate from normality. Thus independence and equality of location parameters are tested by non parametric methods. The first one is tested using Kendall's Tau test; these results suggest that most variables are not independent (see Table \ref{ktau}). Nevertheless, it should be noted that our estimates generally have low values, which suggest the relationships are weak.

\begin{table}[]
\small
\centering
\caption{Kendall's Tau Estimates}
\label{ktau}
\begin{threeparttable}
\begin{tabular}{lccccccc}
& betweenness & closeness & load   & degree  & indegree & outdegree & bus           \\ \hline
betweenness &           & 0.13 * & 0.38 *  & 0.27 *   & 0.27 *    & 0.27 * & 0.31 * \\
closeness   &           &        & -0.01 * & 0.34 *   & 0.34 *    & 0.34 * & 0.07 * \\
load        &           &        &         & 0.17 *   & 0.17 *    & 0.17 * & 0.23 * \\
degree      &           &        &         &          & 0.98 *    & 0.98 * & 0      \\
indegree    &           &        &         &          &           & 0.96 * & 0      \\
outdegree   &           &        &         &          &           &        & 0  
\end{tabular}
\begin{tablenotes}\footnotesize
\item[] $H_0:$Variables are independent.
\item[*] significant at 0.05 significance level.
\end{tablenotes}
\end{threeparttable}
\end{table}

The second one is tested by Mann Whitney test; these results suggest that there are differences between the location parameters of the groups, except in the betweenness centrality (see Table \ref{mannw}). 

\begin{table}[]
\centering
\caption{Mann Whitney Test p-values}
\label{mannw}
\begin{threeparttable}
\begin{tabular}{lcc}
Index       & North vs. South & Chillos vs. Tumbaco \\ \hline
betweenness & 0.29            & 0.06                \\
closeness   & 0.00E+00        & 8.61E-178           \\
load        & 2.35E-104       & 6.68E-07            \\
degree      & 0               & 3.42E-85            \\
indegree    & 0               & 1.16E-94            \\
outdegree   & 0               & 5.07E-96            \\
bus         & 1.18E-31        &                    
\end{tabular}
\begin{tablenotes}\footnotesize
\item[] $H_0:$ The location parameters of the distribution of the variable are the same in each group.
\end{tablenotes}
\end{threeparttable}
\end{table}

\subsection{Spatial distributions}\label{sec.spat_dist}

\begin{figure}
	\centering  	        
    \begin{subfigure}[t]{0.475\textwidth}
		\includegraphics[width=\textwidth]{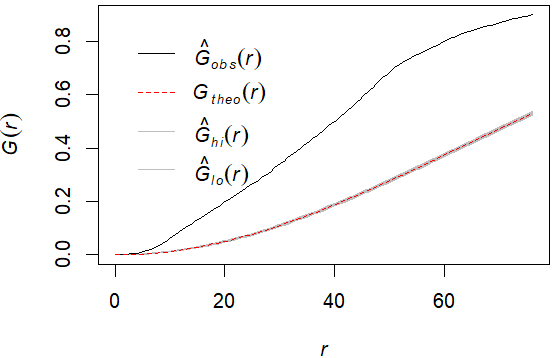}
		\caption{G function}
		\label{fig:Gfunction}
	\end{subfigure}
    ~ \hfill
    \begin{subfigure}[t]{0.475\textwidth}
	\includegraphics[width=\textwidth]{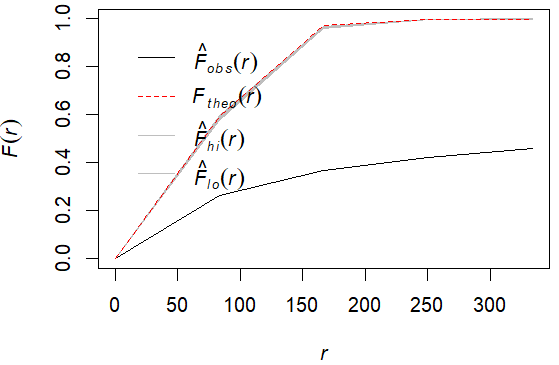}
		\caption{F function}
		\label{fig:Ffunction}
	\end{subfigure}
	\caption{Theoretical and empirical distribution of G and F functions. The gray area around the theoretical distribution (red line) corresponds to the Monte Carlo point wise envelopes.}  
\label{fig:FG}
\end{figure}

After computing node topological measures, a preliminary task to address the spatial distribution of such measures consists on testing the spatial randomness of nodes. In particular, the \textit{Complete Spatial Randomness} hypothesis (CSR) should be tested. This hypothesis states that the nodes are independently and uniformly distributed over the urban map \cite{digglestatistical}. Some methods used for this task are based on the so called \textit{G} and \textit{F} functions. The first one measures the distribution of the distances from an arbitrary event (a node in this case) to its nearest event. The second one measures the distribution of all distances from an arbitrary point of the plane to its nearest event. In both cases, under CSR, the expected value of these functions is given by the following expression, which depends on the neighbourhood radius $r$: $1-exp(-\lambda \pi r^2)$, where, $\lambda$ represents the mean number of events per unit area (or \textit{intensity}). Then point-wise envelopes under CSR are computed by Monte Carlo simulations and compared with the empirical distribution \cite{bivand2008applied}. Figures \ref{fig:Gfunction}) and \ref{fig:Ffunction} show that the empirical function is outside the envelope and takes much larger values than the theoretical function for almost all values of $r$. Hence, the CSR hypothesis is rejected, and we expect that the topological properties of urban network nodes also show a clustered pattern.

This latter idea is confirmed by means of Kernel Density Estimations on the topological measures of nodes. In this case, a cell of 100 meters and a search neighbourhood of 1 kilometer have been considered, and each point (node) is weighted by its corresponding topological feature value. 

\begin{figure}
\centering
\begin{subfigure}[t]{0.45\textwidth}
\centering
\includegraphics[scale=0.7]{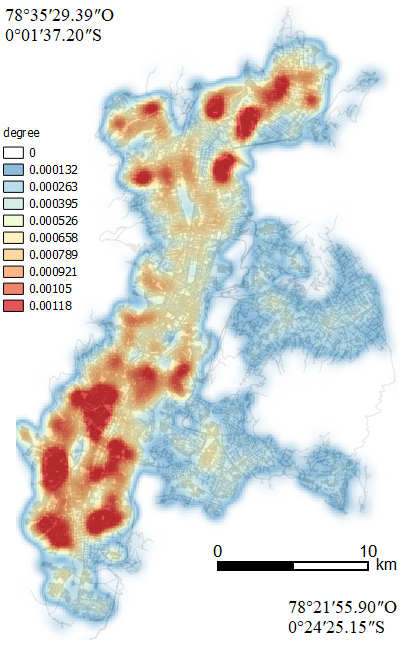}
\caption{}
\label{fig:degree}
\end{subfigure}
~ \hfill
\begin{subfigure}[t]{0.45\textwidth}
\centering
\includegraphics[scale=0.7]{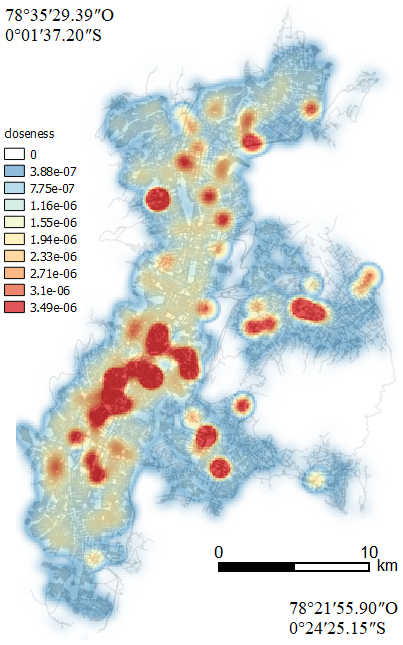}
\caption{}
\label{fig:closeness}
\end{subfigure}
~ \vfill
\begin{subfigure}[t]{0.45\textwidth}
\centering
\includegraphics[scale=0.7]{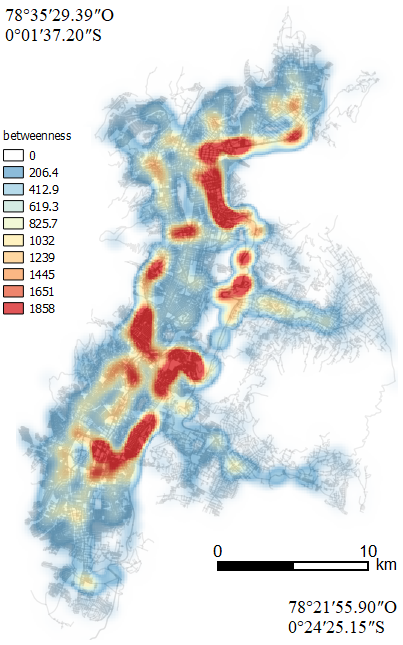}
\caption{}
\label{fig:betweenness}
\end{subfigure}
~ \hfill
\begin{subfigure}[t]{0.45\textwidth}
\centering
\includegraphics[scale=0.7]{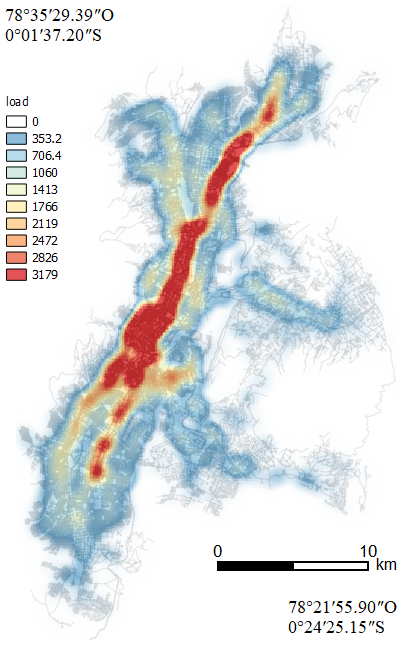}
\caption{}
\label{fig:load}
\end{subfigure}
\caption{Kernel density estimation for node centrality measures.}
\label{fig:kde_centrality}
\end{figure}

At the very first, we considered degree centrality. According to \cite{barthelemy2011spatial}, this index may be not so interesting in spatial networks because of its peaked distribution. Nevertheless, in the case of Quito, it reveals differences in the spatial patterns of the north and the south of the city. In particular, Figure \ref{fig:degree} shows that the majority of nodes with higher (non normalized) degree are clustered in the south of the city. This suggests that this zone has a denser urban network than the other one.

Secondly, we confirm that density estimation gives account of the hierarchical structure of the urban network, i.e., it shows which streets are more important, as in \cite{liu2015road}. According to \cite{ryder2004land}, a prominent feature of Latin American cities is \textit{the spine}, which is described as a single dominant transportation axis where ``commercial and industrial activities are concentrated in a central business district but diffuse outward along such axis''. This author claims that both Avenida Diez de Agosto and Avenida Amazonas act as a \textit{composite spine} of Quito because the first one is the most important for north-south intra-urban transporation, while the second one is attractive for business activities.

Closeness centrality, betweenness centrality, and load centrality hotspot maps (Figures \ref{fig:closeness}, \ref{fig:betweenness} and \ref{fig:load}, respectively) demonstrate that the zones with higher density of topologically important nodes not only lie alongside the spine, but also along other important avenues and highways. It makes sense these highways (e.g. Av. Simón Bolívar) have high centrality values since they border the city and connect the extreme south with the extreme north, and the valleys with the city. A summmary of such important roads in Quito is given in Table \ref{avenues}.

\begin{figure}
\centering
\begin{subfigure}[t]{0.45\textwidth}
\centering
\includegraphics[scale=0.7]{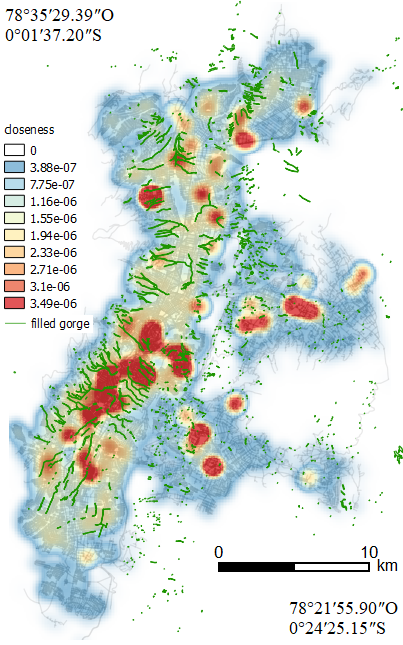}
\caption{}
\label{fig:clos_gorges}
\end{subfigure}
~ \hfill
\begin{subfigure}[t]{0.45\textwidth}
\centering
\includegraphics[scale=0.7]{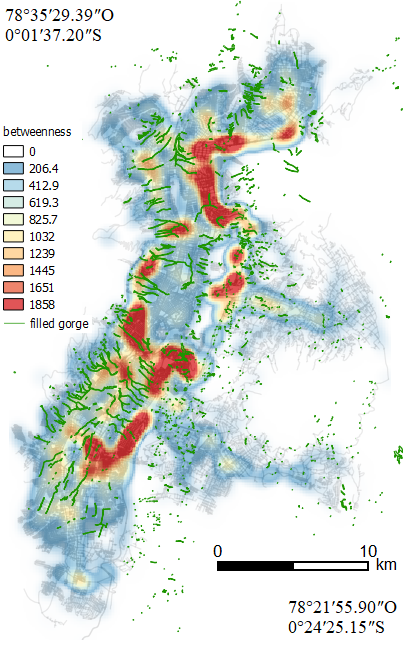}
\caption{}
\label{fig:bet_gorges}
\end{subfigure}
\label{fig:kde_gorges}
\caption{Centrality measures hotspots overlapped with filled gorges (in green).}
\end{figure}

\begin{figure}
	\centering  	        
    \begin{subfigure}[t]{0.475\textwidth}
		\includegraphics[width=\textwidth]{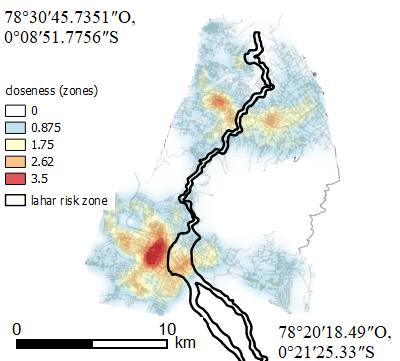}
		\caption{closeness}
		\label{fig:clos_lahar}
	\end{subfigure}
    ~ \hfill
    \begin{subfigure}[t]{0.475\textwidth}
	\includegraphics[width=\textwidth]{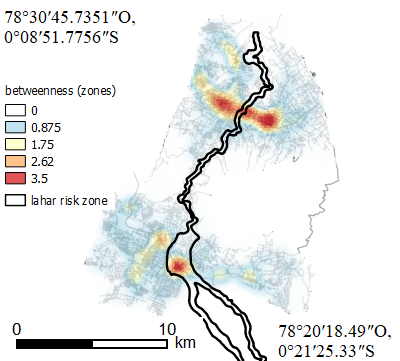}
		\caption{betweenness}
		\label{fig:bet_lahar}
	\end{subfigure}
	\caption{Centrality measures hotspots of the Valleys overlapped with lahar risk zone (in black).}  
\label{fig:lahar}
\end{figure}

Thirdly, some topological index hotspots match with some natural risk zones:

\begin{itemize}
    \item Some sections of the hydrographic network were channeled, and streets were built over filled streams (or gorges). As a result, in the presence of heavy rain the hydrographic network caudal increases, and there is higher risk for sinkholes forming. Since in the next section we will refer to a related event, now we limit ourselves to mentioning that there is a fair overlap between filled gorges and centrality measures hotspots as can be seen in Figures \ref{fig:clos_gorges} and \ref{fig:bet_gorges}. 

    \item Cotopaxi volcano is located in the south of Quito. Some parts of the valleys (Los Chillos and Tumbaco-Cumbayá) are zones of risk if a volcanic eruption occurs. In particular, those zones closer to the rivers Pita, Santa Clara y San Pedro would be most affected by volcanic lahars. In Figure \ref{fig:lahar}, we show that some closeness and betweenness hotspots (computed in the subnetworks of the valleys) match with risks zones for volcanic lahars.

    \item According to \cite{demoraes2005seismic}, those important roads detailed in Table \ref{avenues} contain a fair amount of elevated road infrastructures, which are vulnerable in case of a ``very severe earthquake (PGA = 0.40g)''. 
    
\end{itemize}


\begin{figure}
\centering
\includegraphics[scale=0.65]{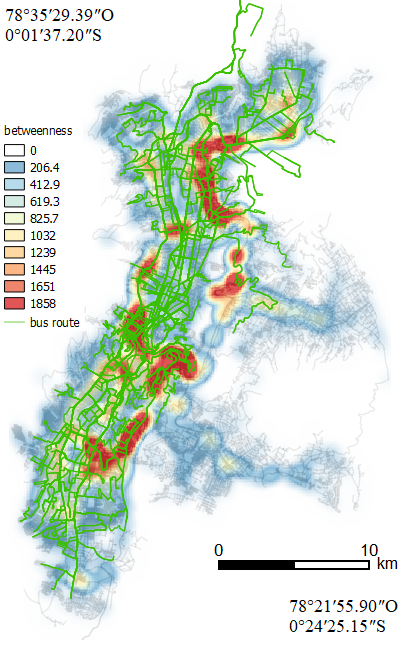}
\caption{Betweenness hotspots overlapped with the principal bus routes (in green).}
\label{fig:bus_bet}
\end{figure}

\begin{figure}
\centering
\includegraphics[scale=0.65]{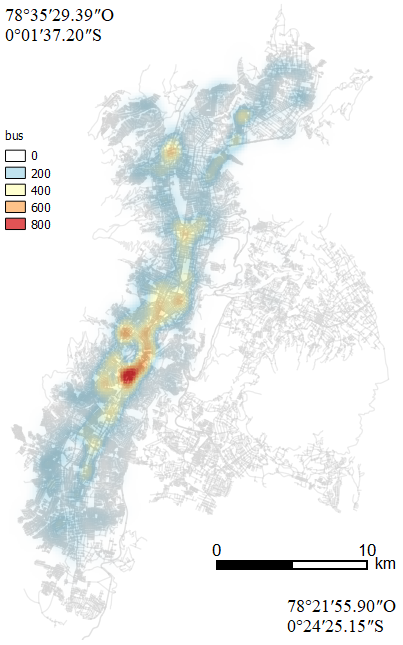}
\caption{Kernel density estimation for bus pass.}
\label{fig:bus}
\end{figure}

Fourthly, we found agreement between topological structure and mobility patterns. These patterns can somehow be addressed by means of the principal bus routes of the city. Figure \ref{fig:bus_bet} shows an overlap between bus routes and betweenness centrality hotspots. Additionally, we computed the number of times a bus route passes through some node and estimated its kernel density. Although data is not fully available for the valleys, we consider the analysis is still worthwhile since such routes are less diverse and coincide with other topological measures that overlap previously mentioned avenues. Figure \ref{fig:bus} shows that there is a zone in the South which is highly trafficked by bus routes, and it is connected to the center of the city. Hotspots in the North are also important since they are in the vicinity of the city's business center.

\begin{figure}
\centering
\begin{subfigure}[t]{0.45\textwidth}
\centering
\includegraphics[scale=0.7]{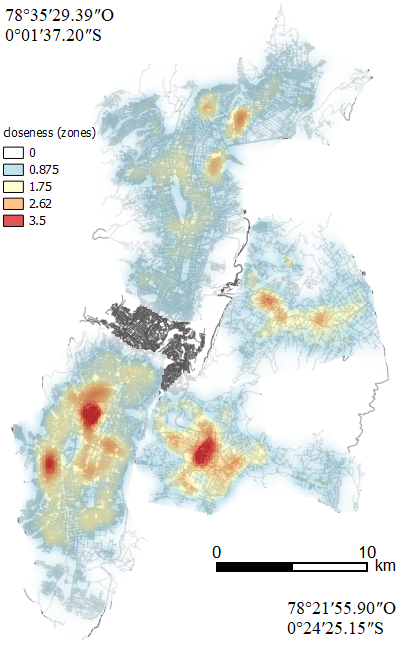}
\caption{}
\label{fig:closenessz}
\end{subfigure}
~ \hfill
\begin{subfigure}[t]{0.475\textwidth}
    \centering
\includegraphics[scale=0.7]{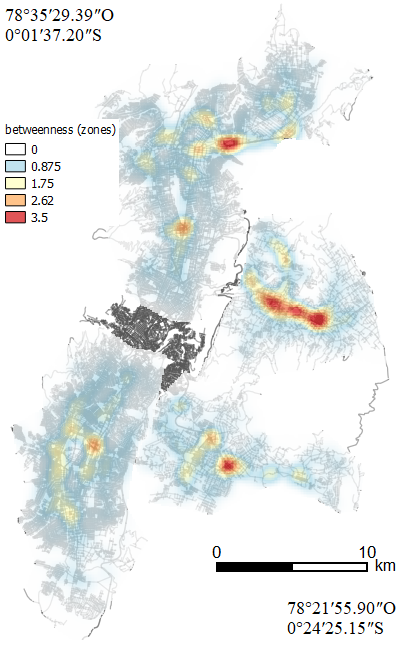}
\caption{}
\label{fig:betweennessz}
\end{subfigure}
~ \hfill
\begin{subfigure}[t]{0.475\textwidth}\centering
\includegraphics[scale=0.7]{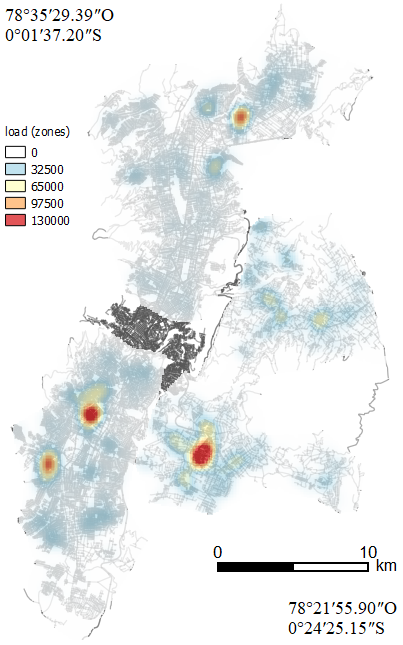}
\caption{}
\label{fig:loadz}
\end{subfigure}
	\caption{Kernel density estimation for centrality measures: Zones of Quito.}  
\label{fig:kde_zones}
\end{figure}

To conclude this subsection, we refer to the structure of each subgraph corresponding to the North, South, Los Chillos, and Tumbaco-Cumbayá zones of Quito. Figures \ref{fig:closenessz}, \ref{fig:betweennessz}, and \ref{fig:loadz} contain each subgraph and the corresponding Kernel Density Estimations.

We note that the historical center appeared much before other zones, so the architectural motivations were different. Additionally, this zone has a subgraph smaller than the other ones. For these reasons, the historical center deserves a separate analysis, which is beyond the scope of this project. Hereinafter, we will compare the North with the South, and the valleys among them due to their similarities in subgraph sizes and antiquity of settlements (the North and South were built way before the valleys). As a note of caution, we mention that the color scales shown in the corresponding figures are the same only for the comparison group.

First, we note that some of the hotspots in the graph of the whole city persist in the subgraphs, i.e., some of the structure of the whole graph is preserved in the subgraphs. This may happen because the elongated shape of the city and its parts do not drastically change the layout of the city. 

Secondly, closeness and load centralities behave similarly. In particular, hotspots of both measures tend to match and they suggest clear differences between zones: higher values tend to cluster more in the South rather than in the North, and in Los Chillos rather than Tumbaco-Cumbayá. On the contrary, betweenness does not suggest large differences between the North and South, except for the hotspot in the north-east which corresponds to the entrance to Quito. Furthermore, Tumbaco-Cumbayá has larger hotspots than Los Chillos. These hotspots  mainly lie on the highway that connects Tumbaco-Cumbayá with the North of the city. 

\subsection{Network Robustness}\label{sec.robustness}

The previous subsection was devoted to identifying the (topologically) important nodes of the city and detect spatial patterns (in particular, hotspots). We have used these results as inputs for assessing the robustness of the urban network robustness under road disruptions. More specifically, we have considered two node removal (attacks) strategies: 1) sequentially removing the most important nodes, as measured by topological centrality and an index based on bus routes crossing; and 2) randomly removing nodes.

\begin{figure}
\centering
\includegraphics[scale=0.35]{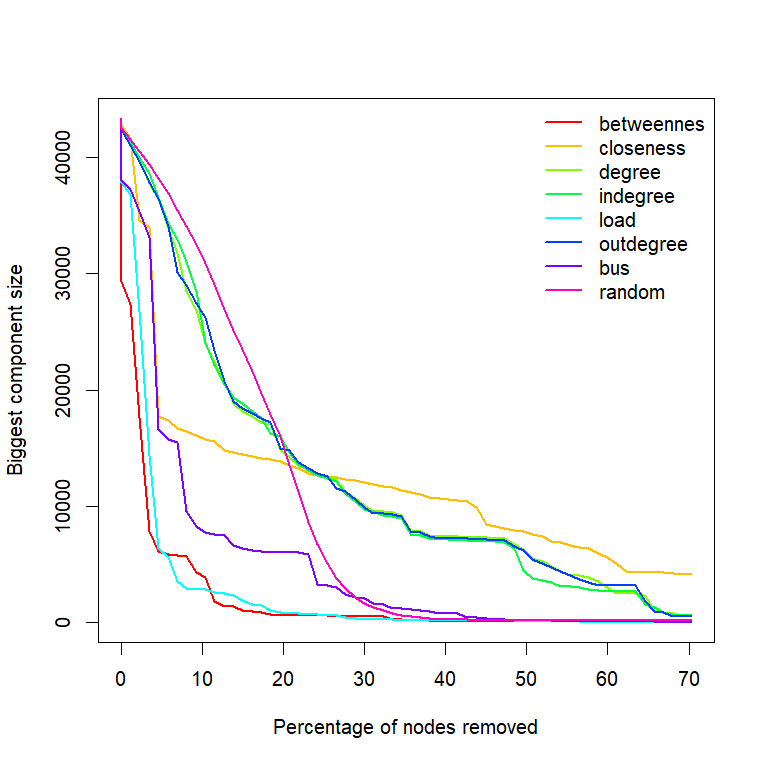}
\caption{Quito Network Robustness under attacks}
\label{fig:attacks}
\end{figure}

Figure \ref{fig:attacks} plots the biggest connected component size as a function of the removed nodes based on some strategies. The observed behaviours show that directed attacks to nodes based on betweenness centrality or load centrality produce the fastest shrinkages in the biggest component size for the whole network. Indeed, only 10$\%$ of nodes are needed to produce a significant affectation in the network. Hence, these nodes, as seen in previous subsections, are important in the extent their failure can interrupt the communication between different parts of the city. A slower but also effective way to disconnect the network consist on performing attacks based on bus routes crossings. Finally, degree and closeness based attacks do not perform much better than random attacks.

It is important to mention that, we have not carried out removals based on sinkholes, lahars, or elevated road infrastructure damage since they are somehow related (or overlapped) with the topological centrality patterns presented in \ref{sec.spat_dist}. Nevertheless, we will refer to them along this subsection for purposes of interpretation.

Regarding sinkholes, we expect that the strategy based on removal of nodes lying on sinkhole risk zones behaves similarly to closeness and betweenness based strategies. In particular, when 10\% of nodes had been disrupted, the size of the biggest connected component would be at most 50\% of the original size. A real event associated with a disruption of this type occurred in 2008, when a part of the road infrastructure (known as El Trébol and built over a filled gorge) sank. It was an important event since it connects the north with south of the city and Los Chillos Valley with the historical center. As a precautionary measure, neighbouring roads were closed, so the mobility was heavily affected. A detailed description of the event and its consequences can be found in \cite{toulkeridis2016causes}.

When considering the robustness of the zones of Quito, we again make comparisons between ancient and mostly urbanized zones (i.e. North and South) and between new and urban developing zones (i.e., Los Chillos and Tumbaco-Cumbayá valleys). 

\begin{figure}
\centering
\begin{subfigure}[t]{0.45\textwidth}
\centering
\includegraphics[scale=0.3]{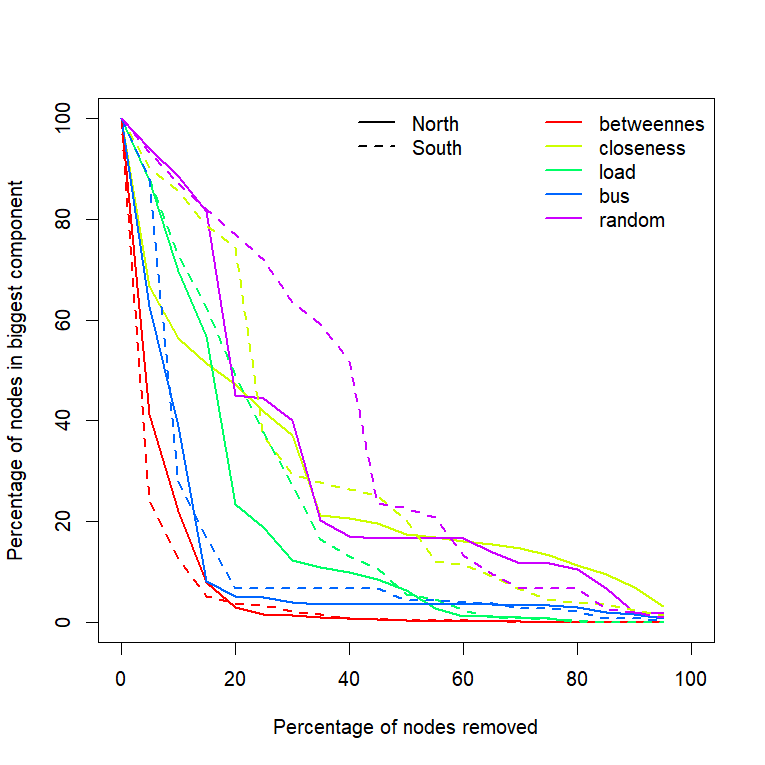}
\caption{}
\label{fig:attacksNS1}
\end{subfigure}
~ \hfill
\begin{subfigure}[t]{0.45\textwidth}
\centering
\includegraphics[scale=0.3]{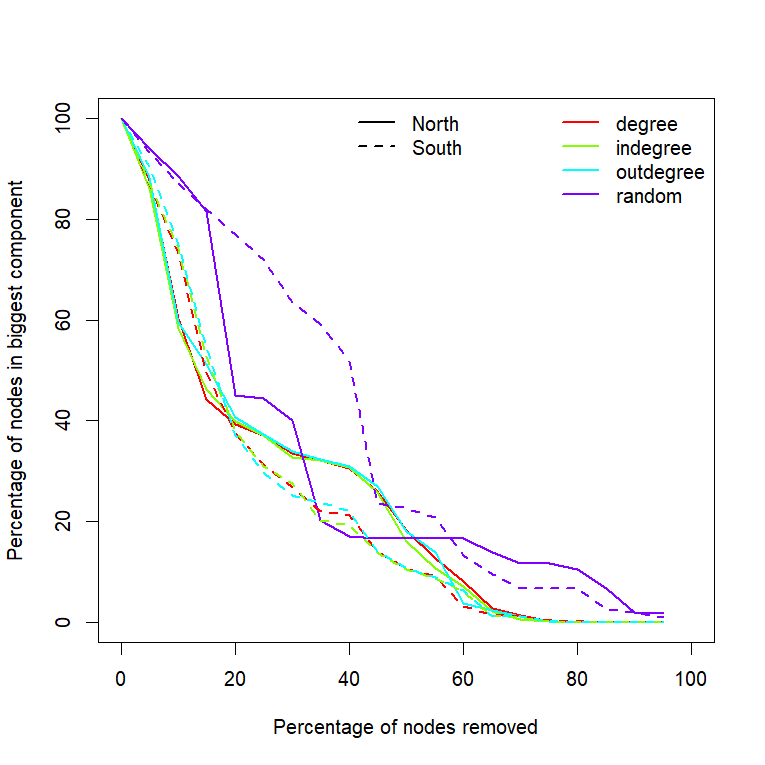}
\caption{}
\label{fig:attacksNS2}
\end{subfigure}
\caption{Network Robustness under attacks: North and South}
\label{fig:attacksNS}
\end{figure}

Regarding the North and South (see Figure \ref{fig:attacksNS1}), it can be noted that betweenness-based and load-based attacks show the same behaviour as in the whole urban network, i.e., they disconnect the networks faster than other strategies. In the first case, the South disconnects slightly faster than the North, while in the second one, the opposite occurs. Furthermore, during almost all the simulated closeness-based attacks, the North disconnects faster than the South. We recall that most of the elevated road infrastructures are located in the North and there is a concentration of facilities in that zone \cite{demoraes2005seismic}. Then disruption of high betweenness nodes, such as those lying on Avenida 10 de Agosto (which supports trolleybus transport system) and Avenida Mariscal Sucre (which is the most traffic loaded road of the city), would severely affect these areas accessibility. On the other hand, degree, indegree, and outdegree exhibit a similar behaviour among them, making it possible to distinguish between two intervals: from $0\%$ to $20\%$ of nodes removed, where the North network decreases faster than the South network; while for greater than $20\%$ of nodes removed, this behaviour reverts itself (see Figure \ref{fig:attacksNS2}).

\begin{figure}
\centering
\begin{subfigure}[t]{0.45\textwidth}
\centering
\includegraphics[scale=0.3]{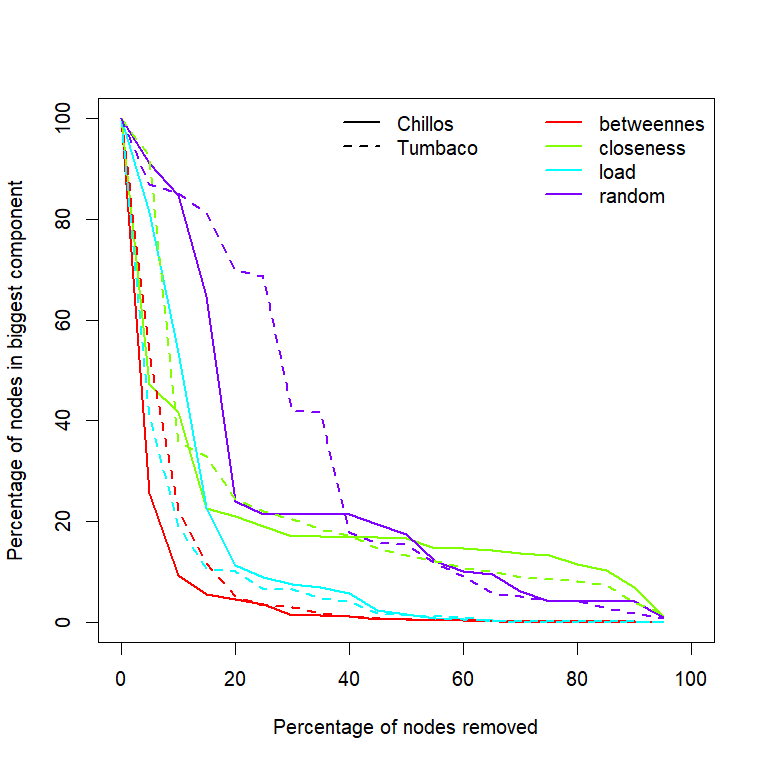}
\caption{}
\label{fig:attacksCT1}
\end{subfigure}
~ \hfill
\begin{subfigure}[t]{0.45\textwidth}
\centering
\includegraphics[scale=0.3]{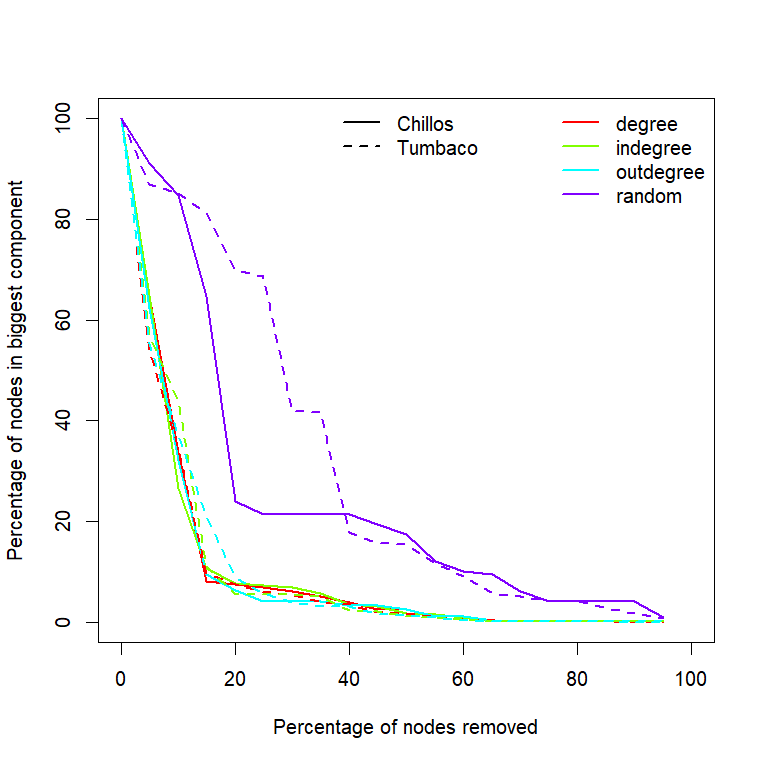}
\caption{}
\label{fig:attacksCT2}
\end{subfigure}
\caption{Network Robustness under attacks: Los Chillos and Tumbaco-Cumbayá}
\label{fig:attacksCT}
\end{figure}

Figure \ref{fig:attacksCT1} and \ref{fig:attacksCT2} show how robust are the networks of the valleys under node removals. At the beginning, it is possible to note that betweenness-based and load-based attacks show the same behaviour as in the whole urban network, i.e., they disconnect the analyzed networks faster than the other strategies. In the first case (betweenness), Los Chillos' biggest connected component shrinks slightly faster than Tumbaco-Cumbayá's connected component, while in the second case, it the opposite occurs. Furthermore, we note that closeness-based attacks (as well as degree, indegree, and outdegree strategies) are not much more effective than random attacks. 

In sec. \ref{sec.spat_dist}, it was shown a match between some closeness and betweenness hotspots (computed in the subnetworks of the valleys) and risk areas for volcanic lahars. Once again, we do not perform  node removals based on such risk areas, but we expect an intermediate behaviour between closeness and betweenness based strategies. Detailed descriptions of Cotopaxi lahars and its potential risks can be found in \cite{Aguilera2004}, \cite{rodriguez2017economic}, and \cite{toulkeridis2018ethics}.

\section{Conclusions}\label{sec.Conclusions}

During the last decade, there has been growing attention for the assessment of urban network vulnerability. Despite the theoretical and methodological progress and the increasingly available data, related work is still sparse in Latin America, even more so in Ecuador. Due to its geographical, historical, and social characteristics (sinkholes, earthquakes, lahars, concentration of facilities, urban sprawl, and others), the city of Quito is exposed to a variety of risks that make it attractive for such an analysis. 

In this context, this project aimed to show some of the potentialities for network analysis in urban planning. In particular, we have explored some spatial patterns that arise in Quito's urban network, and linked them to the aforementioned risks. For this purpose, we have relied upon a methodology consisting on network centrality measures, Kernel Density Estimation, and several strategies of node attacks.

The first important result suggests that high centrality nodes match with critical locations, i.e., places of the city that in case of failure may severely affect the structure and flow in the networks.

The second important result suggests that betweenness and load based strategies are the most effective for disconnecting the urban network. Although the bus route importance index is slightly less effective, it still performs better than other centrality measures.

The third important result suggests that there are spatial differences between the the North and South (the ancient and mostly urbanized zones) and between Los Chillos and Tumbaco-Cumbayá (the new and urban developing zones). Although there are no strong differences in the structure of such networks, which is given by the close behaviour observed when performing node attacks, the inclusion of other aspects in the analysis (e.g., natural risks or concentration of facilities) suggests that there are zones that are more vulnerable than others.

This project has been a first attempt to study the Quito's urban network robustness by graph theoretical methods, so there are several directions that further research may take. 

First, more data can be included for enriching the analysis. For example, socioeconomic data and facility (e.g. hospitals) locations can be considered to test if there are differences between economic classes in the accessibility to those facilities, when road disruptions occur. 

Secondly, cohesive groups can be addressed by means of community structure detection methods. Methods and appropriate null models are still in development for spatial networks. We experimented with some common methods for social networks (e.g. VOS Communities method and Louvain Method) but they produced meaningless results since they are not suitable for weighted and directed multigraphs embedded in space, as urban networks. 

Finally, relations between urban morphology and topology deserve attention since shape patterns may influence the spatial distribution of centrality measures. According to \cite{boeing2017multi}, when referring to US cities, there is a potential relation between shape and topology: a highly connected orthogonal grid is associated with low maximum betweenness centrality. The grid shape is also related with density, because it has been found that US cities with orthogonal grids tend to have high average number of streets per node and low suburban sprawl. Since this has been observed in US cities where flat terrain allows idealized grids, this hypothesis remains to be confirmed in Andean cities, such as Quito, where geography is uneven.

\clearpage

\bibliography{bibliography}{}
\nocite{*}
\bibliographystyle{acm}

\clearpage
\section{Appendix}

\subsection{}
\begin{table}[!ht]
\footnotesize
\centering
\caption{Summary statistics of networks}
\label{sumstat}
\begin{tabular}{lccccc}
Index            & North    & South    & Chillos  & Tumbaco  & Quito    \\ \hline
nodes  & 17251        & 16219        & 3382        & 2963        & 43365        \\
edges  & 42422        & 43660        & 8066        & 6698        & 109159        \\
betweennes.min  & 0        & 0        & 0        & 0        & 0        \\
closeness.min   & 0        & 0        & 0        & 0        & 0        \\
load.min        & 0        & 0        & 0        & 0        & 0        \\
degree.min      & 5.80E-05 & 6.20E-05 & 2.96E-04 & 3.38E-04 & 5.80E-05 \\
indegree.min    & 0        & 0        & 0        & 0        & 0        \\
outdegree.min   & 0        & 0        & 0        & 0        & 0        \\
bus.min         & 0        & 0        &          &          & 0        \\
betweennes.mean & 4.36E-03 & 3.68E-03 & 1.18E-02 & 1.59E-02 & 5.57E-03 \\
closeness.mean  & 1.34E-02 & 1.69E-02 & 2.50E-02 & 2.15E-02 & 1.64E-02 \\
load.mean       & 1.72E+03 & 2.56E+03 & 3.66E+02 & 2.56E+02 & 1.84E+03 \\
degree.mean     & 2.85E-04 & 3.32E-04 & 1.41E-03 & 1.53E-03 & 4.92E-04 \\
indegree.mean   & 1.43E-04 & 1.66E-04 & 7.05E-04 & 7.63E-04 & 2.46E-04 \\
outdegree.mean  & 1.43E-04 & 1.66E-04 & 7.05E-04 & 7.63E-04 & 2.46E-04 \\
bus.median      & 0.00E+00 & 0.00E+00 &          &          & 0.00E+00 \\
betweennes.max  & 3.27E-01 & 1.64E-01 & 3.20E-01 & 3.16E-01 & 3.27E-01 \\
closeness.max   & 2.10E-02 & 2.56E-02 & 3.77E-02 & 3.13E-02 & 3.77E-02 \\
load.max        & 4.01E+04 & 6.88E+04 & 7.47E+03 & 2.57E+03 & 6.88E+04 \\
degree.max      & 6.96E-04 & 7.40E-04 & 3.55E-03 & 3.38E-03 & 3.55E-03 \\
indegree.max    & 3.48E-04 & 3.70E-04 & 1.78E-03 & 1.69E-03 & 1.78E-03 \\
outdegree.max   & 3.48E-04 & 3.70E-04 & 1.78E-03 & 1.69E-03 & 1.78E-03 \\
bus.max         & 2.00E+01 & 2.90E+01 &          &          & 4.00E+01 \\
betweennes.sd   & 1.64E-02 & 1.16E-02 & 3.11E-02 & 3.87E-02 & 1.94E-02 \\
closeness.sd    & 2.76E-03 & 2.98E-03 & 4.91E-03 & 4.37E-03 & 4.76E-03 \\
load.sd         & 2.59E+03 & 3.69E+03 & 5.92E+02 & 3.57E+02 & 3.01E+03 \\
degree.sd       & 1.11E-04 & 1.21E-04 & 6.18E-04 & 6.71E-04 & 5.08E-04 \\
indegree.sd     & 5.64E-05 & 6.07E-05 & 3.10E-04 & 3.38E-04 & 2.54E-04 \\
outdegree.sd    & 5.65E-05 & 6.06E-05 & 3.10E-04 & 3.38E-04 & 2.54E-04
\end{tabular}
\end{table}

\subsection{}

\begin{table}[!ht]
\footnotesize
\centering
\caption{Principal avenues of Quito with high value of centralities}
\label{avenues}
\begin{tabular}{lccc}
Street / Highway             & betweenness & closeness & load \\ \hline
Av. 6 de Diciembre           & x           & x         & x    \\
Av. Eloy Alfaro              & x           & x         & x    \\
Av. Mariscal Sucre           & x           & x         & x    \\
Av. Pedro Vincente Maldonado & x           & x         & x    \\
Panamericana Norte           & x           & x         & x    \\
Av. 10 de Agosto             & x           &           & x    \\
Av. Amazonas                 & x           &           & x    \\
Av. El Inca                  & x           &           & x    \\
Av. Francisco de Orellana    &             & x         & x    \\
Av. Napo                     &             & x         & x    \\
Av. Rodrigo de Chávez        &             & x         & x    \\
Av. Simón Bolívar            & x           &           & x    \\
Av. Velasco Ibarra           &             & x         & x    \\
Av. 12 de Octubre            &             &           & x    \\
Av. Alonso de Angulo         &             &           & x    \\
Av. América                  &             &           & x    \\
Av. Cristóbal Colón          &             &           & x    \\
Av. de la Prensa             &             &           & x    \\
Av. de la República          &             &           & x    \\
Av. Diego de Vásquez         &             & x         &      \\
Av. Galo Plaza Lasso         & x           &           &      \\
Av. General Rumiñahui        &             & x         &      \\
Av. Ilaló                    &             & x         &      \\
Av. Mariana de Jesús         &             &           & x    \\
Av. Morán Valverde           &             & x         &      \\
Av. Oswaldo Guayasamín       &             & x         &      \\
Av. Patria                   &             &           & x    \\
Av. Pichincha                &             &           & x 
\\
Av. Teniente Hugo Ortiz      &             & x         &      \\
Ruta Viva                    &             & x         &      
\end{tabular}
\end{table}

\end{document}